\documentclass[aps,pra, twocolumn,nofootinbib, showpacs]{revtex4-1} 
\usepackage[unicode=true,pdfusetitle, bookmarks=true,bookmarksnumbered=false,bookmarksopen=false, breaklinks=false,pdfborder={0 0 0},backref=false,colorlinks=false] {hyperref}
\hypersetup{ colorlinks,linkcolor=myurlcolor,citecolor=myurlcolor,urlcolor=myurlcolor}
\usepackage{braket,colortbl,amsthm,amsmath,cleveref,amssymb,txfonts}\definecolor{myurlcolor}{rgb}{0,0,0.7}
\usepackage{graphics,graphicx}
\usepackage{color}
\usepackage{amssymb}
\usepackage{amsthm}
\usepackage{amsfonts}
\usepackage{float}
\usepackage{graphicx}
\usepackage[format = plain,labelfont = bf,up, textfont = normal , up, justification =raggedright, singlelinecheck =false]{caption}
\usepackage{subcaption}
\usepackage{tabularx}
\usepackage{amsmath}
\usepackage{braket}

\usepackage{graphicx}
\usepackage[utf8x]{inputenc}
\usepackage{color,soul}
\usepackage{amsmath}
\usepackage{braket}
\usepackage{latexsym}
\usepackage{amssymb}
\usepackage{amsthm}
\usepackage{xcolor}
\usepackage{bm}
\usepackage{graphics,epstopdf}
\usepackage{color}\usepackage{amsmath}
\usepackage{enumitem}
\usepackage{fmtcount}
\usepackage{booktabs}
\usepackage{csquotes}
\usepackage{epsfig}
\usepackage{romannum}
\theoremstyle{plain}

\def\bea{\begin{eqnarray}}
\def\eea{\end{eqnarray}}
\def\ba{\begin{array}}
\def\ea{\end{array}}

\def\ket{\rangle}
\def\bra{\langle}
\def\beq{\begin{equation}}
\def\eeq{\end{equation}}

\begin{document}
\title{Local passivity and entanglement in shared quantum batteries}

\author{Kornikar Sen, Ujjwal Sen}

\affiliation{Harish-Chandra Research Institute, HBNI, Chhatnag Road, Jhunsi, Allahabad 211 019, India}
\begin{abstract}
We identify the conditions for local passivity for
shared quantum batteries with local Hamiltonians. For locally passive
states of two-qubit
batteries, we find the relation of their entanglement content with the
amount of energy that can be globally extracted from them. Moreover,
we obtain that the deficit in work extraction from
pure battery states due to the restriction to local unitaries is equal to
the amount of optimal global work extractable from the corresponding pure
locally
passive battery state, for the same entanglement supply. Furthermore, the
pure battery state for which globally extractable work attains a maximum,
among the set of all pure states with a fixed value of entanglement, also
provides the maximum locally extractable work.
\end{abstract}
\maketitle

\section{Introduction}
\label{sec1}
The amount of energy that is stored in a system and the portion of it that can be extracted is a basic question of thermodynamics, and its utility can hardly be over estimated. An usual battery is one which stores electrochemical energy and converts it to electrical energy when required. These batteries are widely used in various devices. Due to ever-increasing needs of portability and flexibility of devices, batteries of smaller and smaller sizes are required. Hence, creating batteries of molecular size have become a topic of great interest. It may be envisaged that especially the small size will make quantum mechanical effects important in such devices. In recent times, a lot of research is being done in this field and such batteries have been called \enquote{quantum batteries} \cite{Entanglement boost for extractable work from ensembles of quantum batteries, Entanglement Generation is Not Necessary for Optimal Work Extraction, Quantacell: Powerful charging of quantum batteries, Correlation approach to work extraction from finite quantum systems, Enhancing the charging power of quantum batteries, Precision and Work Fluctuations in Gaussian Battery Charging, High-Power Collective Charging of a Solid-State Quantum Battery, Charger-mediated energy transfer in exactly-solvable models for quantum batteries, Spin-chain model of a many-body quantum battery, Quantum Batteries, Bounds on Capacity and Power of Quantum Batteries, Enhanced energy transfer in a Dicke quantum battery, Powerful harmonic charging in a quantum battery, Charger-mediated energy transfer for quantum batteries:
an open system approach, Extractable work the role of correlations and asymptotic freedom in quantum
batteries, Quantum versus classical many-body batteries, Stable adiabatic quantum batteries, Dissipative charging of a quantum battery, Many-body localized quantum batteries, A quantum open system model of molecular battery charged by excitons, Enhancement in performance of quantum battery by ordered and disordered interactions, Charging of quantum batteries with general harmonic power, Random Quantum Batteries, Fluctuations in stored work bound the charging power of quantum batteries, Stabilizing Open Quantum Batteries by Sequential Measurements, Non-Markovian effects on charging of quantum batteries}.

The concept of a quantum battery was, as far as we know, introduced by Alicki and Fannes in 2013 \cite{Entanglement boost for extractable work from ensembles of quantum batteries}. Since then, different models have been considered as substrates for the device, like short- and long-range XXZ quantum spin chains \cite{Spin-chain model of a many-body quantum battery}, spins in cavities modeled by the Dicke interaction \cite{Enhanced energy transfer in a Dicke quantum battery}, ordered and disordered XYZ model \cite{Enhancement in performance of quantum battery by ordered and disordered interactions}, etc. Different methods have been put forward to enhance the charging power of quantum batteries \cite{Quantacell: Powerful charging of quantum batteries, Enhancing the charging power of quantum batteries, Precision and Work Fluctuations in Gaussian
Battery Charging, High-Power Collective Charging of a Solid-State Quantum Battery, Charger-mediated energy transfer in exactly-solvable models for quantum batteries, Powerful harmonic charging in a quantum battery, Charger-mediated energy transfer for quantum batteries:
an open system approach, Dissipative charging of a quantum battery, Charging of quantum batteries with general harmonic power}. The relation between work (i.e., energy) extraction or charging power and entanglement among the batteries when working with more than one battery has been an area of vigorous study \cite{Entanglement boost for extractable work from ensembles of quantum batteries, Correlation approach to work extraction from finite quantum systems, Enhanced energy transfer in a Dicke quantum battery, Quantacell: Powerful charging of quantum batteries, Enhancing the charging power of quantum batteries, High-Power Collective Charging of a Solid-State Quantum Battery, Entanglement Generation is Not Necessary for Optimal Work Extraction}.

 Formally, a quantum battery is a quantum mechanical system described by a state, say $\rho$, and a Hamiltonian, say $H$. One can charge the system by applying a time-dependent field, the system is henceforth assumed to store the energy, and then one can extract work from it by using another time-dependent field. Let the time-dependent field be applied from time $0$ to say $\tau$, for extracting energy.  Then the amount of extracted work is given by 
\begin{equation}
W=\text{Tr} \left(\rho H\right)-\text{Tr} \left(U(\tau, \xi)\rho U(\tau, \xi)^\dagger H\right). \nonumber
\end{equation}
Here, $\xi$ represents the collection of all system parameters, contained, e.g., in the system's potential energy. The dependence of $U$ on $\tau$ and $\xi$ will henceforth be suppressed in the notation. This extracted work will be maximized if we get a unitary operator for which the second term gets a minimum value. Hence, this maximum value of $W$ is given by
\begin{equation}
W_{max}=\text{Tr} \left(\rho H\right)-\min_U \text{Tr} \left(U\rho U^\dagger H\right). \nonumber
\end{equation}
A state from which work extraction is not possible is called a \enquote{passive state} \cite{passive1, passive2}. A passive state, say $\sigma$, of a system with a fixed Hamiltonian, commutes with the Hamiltonian, and if the energy eigenvalues, $\epsilon_i$, satisfy $\epsilon_i<\epsilon_j$, then eigenvalues, $p_i$, of the passive state satisfy $p_j\leq p_i$ for all eigenvalues. The maximum amount of extractable work, in terms of the passive state, is 
\begin{equation}
W_{max}=\text{Tr} \left(\rho H\right)-\text{Tr} \left(\sigma_\rho H\right), \label{eq1}
\end{equation}
where $\sigma_\rho$ is the passive state of the system with Hamiltonian $H$ and has the same eigenvalues as those of the initial state $\rho$. Such a passive state is unique.

In this paper, we define a locally passive state as one from which no
energy can be extracted by using local unitary operations. We provide a
characterization of the same, and prove its uniqueness. We subsequently
restrict attention to two-qubit batteries, and first uncover the relation
between globally extractable work from locally passive battery states and
the entanglement content of the latter. We then consider the issue of
global extraction of work from generic states - not necessarily locally
passive. We identify that the difference between global and local
extraction of work from a pure battery state with a given amount of
entanglement is exactly equal to the optimal global work that is
extractable from the corresponding locally passive battery state having
the same entanglement. Furthermore, we also find that the pure battery
state for which globally extractable work attains a maximum, among the set
of all pure states with a fixed value of entanglement, also
provides the maximum locally extractable work.

We uncover the structure and properties of the locally passive state in
Sec. \ref{sec2}. We present the maximum amount of global work extraction from
these local passive states as a function of their entanglement in Sec.
 \ref{sec3}. In Sec.  \ref{sec4}, we discuss about the maximum global work extraction from
arbitrary states with fixed entanglement. We compare the maximum work
extraction by global and local unitary operations in Sec. \ref{sec5}. We summarize
our results in Sec.  \ref{sec6}.


     
\section{Construction of locally passive quantum battery states}
\label{sec2}
In this section, we discuss about work extraction using \emph{local} unitary operations. If we consider a Hamiltonian, $H_{AB}$, and a system state, $\rho_{AB}$ on a Hilbert space $\mathcal{H}_A\otimes \mathcal{H}_B$, where $A$ and $B$ are two subsystems, then the initial energy of the system is Tr$\left(\rho_{AB} H_{AB}\right)$, and after a local unitary operation, $U_A\otimes U_B$, the energy would be Tr$\left(U_A\otimes U_B \rho_{AB} U_A^\dagger\otimes U_B^\dagger H_{AB}\right)$. Hence, the maximum work extraction using local unitary operations is given by
\begin{equation}
W_{max}^l=\text{Tr} \left(\rho_{AB} H_{AB}\right)-\min_{U_A\otimes U_B}\text{Tr} \left(U_A\otimes U_B \rho_{AB} U_A^\dagger\otimes U_B^\dagger H_{AB}\right). \nonumber \label{eq16} 
\end{equation}
Now, a state $\sigma^l_{AB}$ will be refereed to as  \enquote{locally passive}, if no work can be extracted from this state locally. That is, if Tr$\left(U_A\otimes U_B \sigma_{AB}^l U_A^\dagger\otimes U_B^\dagger H_{AB}\right)\geq  \text{Tr} \left(\sigma_{AB}^l H_{AB}\right)$ for all $U_A\otimes U_B$. Therefore, maximum amount of locally extractable work is given by
\begin{equation}
W_{max}^l=\text{Tr} \left(\rho_{AB} H_{AB}\right)-\text{Tr} \left(\sigma^l_{\rho_{AB}} H_{AB}\right) \label{eqB} .
\end{equation}  
Here $\sigma^l_{\rho_{AB}}$ is a locally passive state of the system with Hamiltonian $H_{AB}$ and has the  same eigenvalues as $\rho_{AB}$.

To uncover some properties of this state, we state the following two theorems. The symbol $I$, with a suitable suffix, denotes the identity operator on the corresponding Hilbert space. \vspace{0.5cm} \\
\noindent \textbf{Theorem 1.} \textit{For self adjoint operators $X_{AB}=X_A\otimes I_B+I_A\otimes X_B$ and $Y_{AB}$ on a finite dimensional Hilbert space $\mathcal{H}_{AB}=\mathcal{H}_A\otimes\mathcal{H}_B$,} Tr\textit{$\left(U_A\otimes U_B Y_{AB} U_A^\dagger\otimes U_B^\dagger X_{AB}\right)\geq$} Tr\textit{$(Y_{AB} X_{AB})$ is true, for all unitary operators $U_A$ and $U_B$, if and only if $X_{A/B}$ commutes with $Y_{A/B}$, where} $\text{Tr}_{A/B}\left(Y_{AB}\right)=Y_{B/A}$\textit{, and eigenvalues of $X_{A/B}$ (say, $\epsilon_i^{A/B}$) and of $Y_{A/B}$ (say $\alpha_i^{A/B}$) satisfy $\left(\alpha^{A/B}_j-\alpha^{A/B}_k)(\epsilon^{A/B}_j-\epsilon^{A/B}_k\right)\leq 0$ for all $j$ and $k$.} \\
\textbf{Remark:} After the statements and proofs of the two theorems, we will identify $X_{AB}$ as a local Hamiltonian $H_{AB}$, and $Y_{AB}$ as a local passive state $\sigma^l_{AB}$. \\
\textbf{Proof.}  \\
Let $X_{A}$ and $X_B$ respectively commute with $Y_{A}$ and $Y_B$, and have common eigenbases $|e_i\ket$ and $|f_i\ket$ respectively. Then in this eigenbases, $Y_{AB}$ and $X_{AB}$ are given by 
\begin{eqnarray}
Y_{AB}&=&\sum a^{kl}_{ij}|e_i\ket\bra e_j|\otimes|f_k\ket\bra f_l|, \label{eq2} \\
X_{AB}&=&\sum\left(\epsilon_m^A|e_m\ket \bra e_m|\otimes |f_n\ket \bra f_n|+\epsilon_n^B|e_m\ket \bra e_m|\otimes |f_n\ket \bra f_n| \right). \label{eq3}
\end{eqnarray}
In this paper, we use the convention in which if the running variable is not mentioned below the summation symbol, then all running variables on the right of it are to be summed over. 
Here, the sums are over the whole eigenbases, and $\epsilon^A_m$ and $\epsilon^B_m$ are eigenvalues of $X_A$ and $X_B$ respectively. Hence, using Eqs. \eqref{eq2} and \eqref{eq3} we get
\begin{eqnarray}
&\text{Tr}&\left(U_A\otimes U_B Y_{AB} U_A^\dagger\otimes U_B^\dagger X_{AB}\right) \nonumber \\
&=&\sum(U_A)_{mi}(U_B)_{nk} a^{kl}_{ij}(U_A^\dagger)_{jm}(U_B^\dagger)_{ln}\epsilon_m^A \nonumber \\
&&+\sum(U_A)_{mi}(U_B)_{nk} a^{kl}_{ij}(U_A^\dagger)_{jm}(U_B^\dagger)_{ln}\epsilon_n^B, \nonumber 
\end{eqnarray}
where the trace is performed in the product bi-orthonormal basis, $\{|e_i\ket \otimes |f_j\ket\}$.
Now, for any unitary operator $U$, we know that $\sum_\beta(U)_{\alpha\beta}(U^\dagger)_{\beta\gamma}=\delta_{\alpha\gamma}$. From Eq. \eqref{eq2}, we can write $Y_A=\sum a^{kk}_{ij}|e_i\ket\bra e_j|$ and $Y_B=\sum a^{kl}_{ii}|f_k\ket\bra f_l|$. Since $Y_A$ and $Y_B$ are represented in terms of their own eigenbases, we have $\sum_ka^{kk}_{ij}=\delta_{ij}\alpha_i^A$ and $\sum_ia^{kl}_{ii}=\delta_{kl}\alpha_k^B$. Using these relations in the above equation, we get
\begin{eqnarray}
&\text{Tr}&\left(U_A\otimes U_B Y_{AB} U_A^\dagger\otimes U_B^\dagger X_{AB}\right) \nonumber \\
&=&\sum \left(|\left(U_A\right)_{mi}|^2\epsilon_m^A\alpha_i^A+|\left(U_B\right)_{nk}|^2\epsilon_n^B\alpha_k^B\right). \nonumber
\end{eqnarray}
Now, $|\left(U_A\right)_{mi}|^2$ and $|\left(U_B\right)_{nk}|^2$ are doubly stochastic matrices, i.e., $\sum_p|\left(U_{A/B}\right)_{pq}|^2=\sum_q|\left(U_{A/B}\right)_{pq}|^2=1$ \cite{Birkhoff}. Therefore, using Birkhoff theorem, we can write $|U_{A/B}| = \sum_r \theta^{A/B}_r P^{A/B}_r$, where $\sum_r\theta_r=1$, $\theta_r\geq 0$ for all $r$, and $P_r$ are permutation matrices of the same Hilbert space. Hence, we get
\begin{eqnarray}
&\text{Tr}&\left(U_A\otimes U_B \rho_{AB} U_A^\dagger\otimes U_B^\dagger H_{AB}\right) \nonumber \\
&=& \sum_r \theta_r^A\sum_m \epsilon_m^A\alpha^A_{r(m)}+\sum_r \theta_r^B\sum_m \epsilon_m^B\alpha^B_{r(m)}. \nonumber
\end{eqnarray}
Here $r$ denotes different permutations. We can see that if $\epsilon^{w}_m>\epsilon^{w}_i$ implies $\alpha^w_m\leq\alpha^w_i$ for both $w=A,B$, then minimum value of the above expression is $\sum \epsilon_m^A\alpha^A_m+\sum \epsilon_n^B\alpha^B_n$, i.e., Tr$(X_{AB}Y_{AB})$. Hence, we get Tr$\left(U_A\otimes U_B Y_{AB} U_A^\dagger\otimes U_B^\dagger X_{AB}\right)\geq \text{Tr}\left(Y_{AB}X_{AB}\right)$. Thus one part of the theorem is proved.

Let us next assume that Tr$\left(U_A\otimes U_B Y_{AB} U_A^\dagger\otimes U_B^\dagger X_{AB}\right)\geq \text{Tr}\left(Y_{AB}X_{AB}\right)$ is true for any $U_A\otimes U_B$. Now if we expand $U_A$ and $U_B$ as 
\begin{eqnarray}
U_A=1+2M_A+2M_A^2+2M_A^3+\cdot\cdot\cdot, \nonumber \\
U_B=1+2M_B+2M_B^2+2M_B^3+\cdot\cdot\cdot,  \nonumber 
\end{eqnarray}
where $M^\dagger_{A/B}=-M_{A/B}$ and $||M_{A/B}||<1$, then using these and simplifying a little bit, we get
\begin{eqnarray}
&\text{Tr}&\left(U_A\otimes U_B Y_{AB} U_A^\dagger\otimes U_B^\dagger X_{AB}\right) \nonumber \\
&=&\text{Tr}\left(X_{AB}Y_{AB}\right)+2\text{Tr}\left(\left[Y_A,X_A\right]M_B\right)+2\text{Tr}\left(\left[Y_B,X_B\right]M_A\right). \nonumber
\end{eqnarray}
From the above equation, we can see that the minimum value of either side would be at $U_A=I_A$ and $U_B=I_B$, where $I_{A/B}$ is the identity operator in the Hilbert space $\mathcal{H}_{A/B}$, if $[Y_{A/B},X_{A/B}]=0$.  
Now, let the unitaries $U_A$ and $U_B$ in the subspace of eigenvectors corresponding to any two eigenvalues $\epsilon^A_1$, $\epsilon^A_2$ of $X_A$ and $\epsilon^B_1$, $\epsilon^B_2$ of $X_B$ be given by

\begin{eqnarray}
U_A^s=
\left[
\begin{matrix}
\cos\phi_A&\sin\phi_A \\
-\sin\phi_A&\cos\phi_A
\end{matrix}
\right],
\ \ U_B^s=
\left[
\begin{matrix}
\cos\phi_B&\sin\phi_B \\
-\sin\phi_B&\cos\phi_B
\end{matrix}
 \right]. \nonumber
\end{eqnarray}
Suppose that in this subspace, $Y_{AB}$ is given by
\begin{equation}
Y_{AB}^s=
\left[
\begin{matrix}
a_1&a_2&a_3&a_4 \\
b_1&b_2&b_3&-a_3 \\
c_1&c_2&c_3&-a_2 \\
d_1&-c_1&-b_1&d_4
\end{matrix}
\right]. \nonumber
\end{equation}
Hence, 
\begin{eqnarray}
Y_A^s=
\left[
\begin{matrix}
a_1+b_2&0\\
0&c_3+d_4

\end{matrix}
\right], \ \ 
Y_B^s=
\left[
\begin{matrix}
a_1+c_3&0\\
0&b_2+d_4
\end{matrix}
\right]. \nonumber
\end{eqnarray} 
These are diagonal matrices, with eigenvalues $a_1+b_2$ (say $\alpha_1^A$), $c_3+d_4$ (say $\alpha_2^A$) and  $a_1+c_3$ (say $\alpha_1^B$), $b_2+d_4$ (say $\alpha_2^B$), as they should be, because we have written all the matrices in the eigenbases of $X_A$ and $X_B$. Using these matrices, we get
\begin{eqnarray}
&&\text{Tr}\left({U^s_A}\otimes {U^s_B} Y_{AB}^s {U^s_A}^\dagger\otimes {U^s_B}^\dagger X^s_{AB}\right) \nonumber \\
&=&\frac{1}{2}\left[\left(\alpha_1^A+ \alpha_2^A\right) \left(\epsilon_1^A + \epsilon_2^A + \epsilon_1^B + \epsilon_2^B\right)\right] \nonumber \\ 
&+&\frac{1}{2}\left[\left(\alpha_1^A-\alpha_2^A\right) \left(\epsilon_1^A - \epsilon_2^A\right) \cos(2\phi_A)\right] \nonumber \\ 
&+& \frac{1}{2}\left[\left(\alpha_1^B-\alpha_2^B\right) \left(\epsilon_1^B - \epsilon_2^B\right) \cos(2 \phi_B)\right], \nonumber
\end{eqnarray}
where $X^s_{AB}$ is $X_{AB}$ in the corresponding subspace. This would be minimum at $\phi_A=0$ and $\phi_B=0$ only if $\left(\epsilon_1^A - \epsilon_2^A\right)>0$ for $ \left(\alpha_1^A -\alpha_2^A\right)\leq0$ and $\left(\epsilon_1^B - \epsilon_2^B\right)>0$ for $\left(\alpha_1^B -\alpha_2^B\right)\leq 0$. \hfill \(\square\)
\vspace{0.5cm} \\
\noindent\textbf{Theorem 2.}
\textit{Every system has a unique locally passive state.} \\
\textbf{Proof.} \\
Let us begin by assuming that the converse statement is true, i.e., a system $\rho_{AB}$ has two locally passive states, $\sigma^l_{\rho_{AB}}$ and $\sigma'^l_{\rho_{AB}}$, which are related to $\rho_{AB}$ through local unitary operations. Hence, $\sigma^l_{\rho_{AB}}$ and $\sigma'^l_{\rho_{AB}}$ are also related through a local unitary operation, say $U_A \otimes U_B$, where 
\begin{eqnarray}
U_w^s=\left[
\begin{matrix}
\cos\phi_w&\sin\phi_w \\
-\sin\phi_w&\cos\phi_w
\end{matrix} 
\right]. \nonumber
\end{eqnarray}
Here, $w$ denotes either of the two subsystems, $A$ and $B$, and the superscript $s$ indicates that we are working in a two-dimensional subspace (see proof of Theorem 1). Now, let
\begin{eqnarray}
 \text{Tr}_B\left(\sigma^{ls}_{\rho_{AB}}\right)= 
 \left[
 \begin{matrix}
 p&0\\
 0&q
 \end{matrix}
 \right]. \nonumber
\end{eqnarray} 
After applying $U_A$ on $ \text{Tr}_B\left(\sigma^{l}_{\rho_{AB}}\right)$, we would get $ \text{Tr}_B\left(\sigma'^{l}_{\rho_{AB}}\right)$. Since $\text{Tr}_B\left(\sigma'^{l}_{\rho_{AB}}\right)$ is another passive state in the subsystem $A$, it should also be a diagonal matrix, in the same basis in which $\text{Tr}_B\left(\sigma^{l}_{\rho_{AB}}\right)$ is expressed above. Hence the off-diagonal term, $(p-q)\sin\phi_A \cos\phi_A=0$. Therefore, $\phi_A=0$, $\frac{\pi}{2}$, $\pi$, $\frac{3\pi}{2}$. Similarly, $\phi_B=0$, $\frac{\pi}{2}$, $\pi$, $\frac{3\pi}{2}$. For $\phi_{A/B}=0$, $\pi$, we will get the same state back, and for the other two values of $\phi_{A/B}$ we will get a state with the same eigenvalues but in opposite order, so that it would not be a local passive state. Hence a system would always have an unique locally passive state. \hfill \(\square\)

From the two theorems above, we conclude that if a system, in the state $\sigma_{AB}^l$, and governed by the Hamiltonian $H_{AB}=H_A\otimes I_B+I_A\otimes H_B$, satisfies the conditions 
\begin{enumerate}
\item subsystems of $\sigma_{AB}^l$, i.e. $\sigma^l_A$ and $\sigma^l_B$ commutes with $H_A$ and $H_B$, and \label{con1} 
\item if eigenvalues of $H_A$ and $H_B$ are set in increasing order, then in the corresponding basis, eigenvalues of $\sigma_{A}^l$ and $\sigma_B^l$ would be in non-increasing order, \label{con2}
\end{enumerate}
then one cannot extract any work from this state by local unitary operations. Hence, these are the locally passive states. For any system $\rho_{AB}$ and Hamiltonian $H_{AB}=H_A\otimes I_B+I_A\otimes H_B$, we can get a corresponding local passive state $\sigma_{\rho_{AB}}^l$ such that 
\begin{equation}
\sigma_{\rho_{AB}}^l=U_A\otimes U_B \rho_{AB} U_A^\dagger \otimes U_B^\dagger , \nonumber
\end{equation} 
where $U_A$ and $U_B$ are the unitaries which diagonalizes $\rho_A$ and $\rho_B$ in such a way that its eigenvalues are non-increasing with eigenvalues of $H_A$ and $H_B$. This therefore forms a complete characterization of the locally passive states for local Hamiltonian, in arbitrary dimensions.
\section{Global work extraction from locally passive quantum battery states with fixed entanglement}
\label{sec3}
In this section, we will consider the problem of global work extraction for two-qubit states which are locally passive for local Hamiltonians, where the state has a pre-decided amount of entanglement shared. We begin our analysis with pure states, and then generalize to mixed states. We will henceforth do all calculations by considering the Hamiltonian, 
\begin{equation}
H_{AB}=\epsilon^A\sigma_z\otimes I_2+\epsilon^B I_2\otimes\sigma_z, \label{eq4}
\end{equation}
of a two-qubit system, where $\epsilon^A>\epsilon^B$, $\sigma_z$ is the Pauli $z$ matrix, and $I_2$ is the identity operator on the qubit Hilbert space. Note that this amounts to choosing a local basis at the outset for a two-qubit Hamiltonian of the form $H_A\otimes I_A +I_B\otimes H_B$, and hence does not lead to any loss of generality for our purposes. We also assume that $\epsilon_A > \epsilon_B\geq 0$. Whenever we will do any numerical calculation we will, for specificity, take $\epsilon^A=2\epsilon$ and $\epsilon^B=\epsilon$, where $\epsilon$ has the units of energy. Now, in this basis, energy eigenvalues are in decreasing order, if chosen as per the sequence, $\epsilon^A+\epsilon^B$, $\epsilon^A-\epsilon^B$, $-\epsilon^A+\epsilon^B$, $-\epsilon^A-\epsilon^B$. A general pure locally passive state is given by 
\begin{eqnarray}
\sigma_{AB}^l=
\left[
\begin{matrix}
|c_0|^2&c_0c_1^*&c_0c_2^*&c_0c_3^* \\
c_1c_0^*&|c_1|^2&c_1c_2^*&c_1c_3^*  \\
c_2c_0^*&c_2c_1^*&|c_2|^2&c_2c_3^*  \\
c_3c_0^*&c_3c_1^*&c_3c_2^*&|c_3|^2 
\end{matrix}
\right], \label{eq5} 
\end{eqnarray}
where the $c_i$'s satisfy the following conditions:
\begin{enumerate}

\item[(i)] $|c_i|\leq 1$ for $i=0,1,2,3$, 
\item[(ii)] $\sum_i |c_i|^2=1$, 
\item[(iii)] $c_0c_2^*=-c_1c_3^*$, 
\item[(iv)] $c_0c_1^*=-c_2c_3^*$,
\item[(v)] $|c_0|^2+|c_1|^2\leq |c_2|^2+|c_3|^2$,  
\item[(vi)] $|c_0|^2+|c_2|^2\leq |c_1|^2+|c_3|^2$. 
\end{enumerate}
Here, the first two conditions ensure that the state is a valid pure quantum state, while the next four ensure its local passivity. In this paper, we will measure entanglement by using the concept of logarithmic negativity \cite{measure}. The amount of entanglement in $\sigma^l_{AB}$ is given by 
\begin{equation}
E=\log_2\left(2|c_1c_2-c_0c_3|+1\right). \label{eq6} 
\end{equation}
Since $\sigma_{AB}^l$ is a pure state, its eigenvalues are 1, 0, 0, 0.
Hence, using Eq. \eqref{eq1}, we get the maximum work that is extractable by using global unitary operations   from $\sigma^l_{AB}$, and is given by
\begin{eqnarray}
W_{max}^p=\left(\epsilon^A+\epsilon^B\right)\left(|c_0|^2-|c_3|^2+1\right)+\left(\epsilon^A-\epsilon^B\right)\left(|c_1|^2-|c_2|^2\right). \nonumber \\ \label{eq7}
\end{eqnarray}
The superscript \enquote{$p$} is to indicate that the state from which the work extraction is being considered is locally passive. If we substitute the values of $c_0$ and $c_1$ from condition (iii) in condition (iv), we get the following results:
\begin{enumerate}
\item[(a)] If $c_3=0$ but $c_2\neq 0$, then $c_0=0$. Hence, $W^p_{max}=\left(\epsilon^A-\epsilon^B\right)\left(|c_1|^2-|c_2|^2\right)+\epsilon^A+\epsilon^B$. \label{con8}
\item[(b)] If $c_2=0$ but $c_3\neq 0$, then $c_1=0$. \\ Hence, $W^p_{max}=\left(\epsilon^A+\epsilon^B\right)\left(|c_0|^2-|c_3|^2+1\right)$. \label{con9}
\item[(c)] If $c_3\neq 0$ and $c_2\neq 0$, then $|c_1|^2=|c_2|^2$ and $|c_3|^2=|c_0|^2$.\\ Hence, $W^p_{max}=\epsilon^A+\epsilon^B$. \label{con10}
\end{enumerate}
We can see that $W^p_{max}$ has higher values for case (b) than in the other two. Hence among all pure local passive states, we can extract higher work from those states for which  $c_1=c_2=0$. Putting $c_1=c_2=0$ in Eq. \eqref{eq6}, and expressing $c_0$ and $c_3$ in terms of $E$, using condition (ii) and Eq. \eqref{eq6}, we get
\begin{eqnarray}
|c_0|&=&\frac{1}{2}\left(1\pm\sqrt{1-(2^E-1)^2}\right) \nonumber \\
\text{and } |c_3|&=&\frac{1}{2}\left(1\mp\sqrt{1-(2^E-1)^2}\right). \label{eq8}
\end{eqnarray}
But according to conditions (v) and (vi), $|c_0|\leq |c_3|$, and hence the acceptable solution is $|c_0|=\frac{1}{2}\left(1-\sqrt{1-(2^E-1)^2}\right)$.
Thus, the maximum amount of extractable work from pure locally passive states with a fixed entanglement $E$, is \begin{equation}
G_E^p=(\epsilon^A+\epsilon^B)\left(1-\sqrt{2^{E+1}-2^{2E}}\right). \label{A}
\end{equation}
We denote the state, for which this maximum value is achievable, by $\sigma^{l\text{ } max}_E$.
\begin{figure}[h!]
\includegraphics[scale=.8]{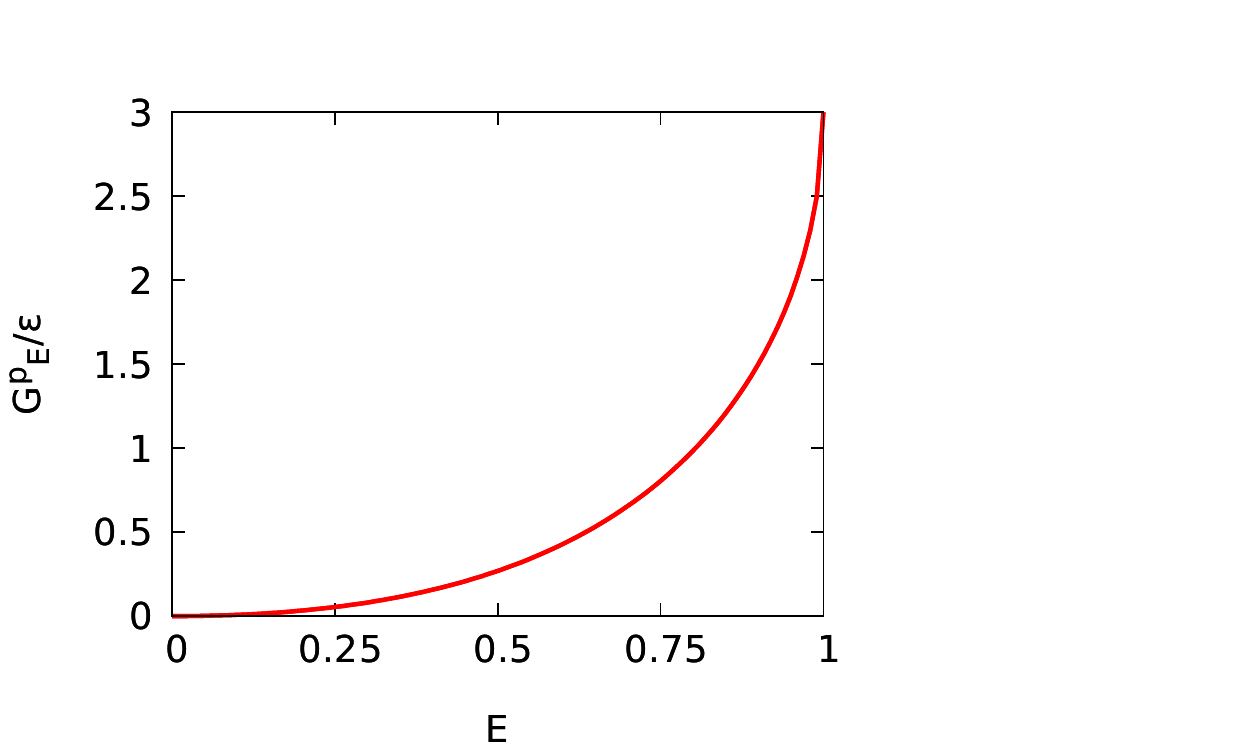}
\caption{Global work extractable from pure locally passive quantum battery states. The maximum work extractable from a set of pure locally passive state with fixed entanglement E, using global unitary operations is denoted by $G^p_E$, and is plotted in units of $\epsilon$, on the vertical axis, while $E$ is plotted on the horizontal one. The horizontal axis is in ebits, while the vertical one is dimensionless. The analytic form is given in Eq. \eqref{A}.} 
\label{fig1}
\end{figure}
\begin{figure}[h!]
\includegraphics[scale=.8]{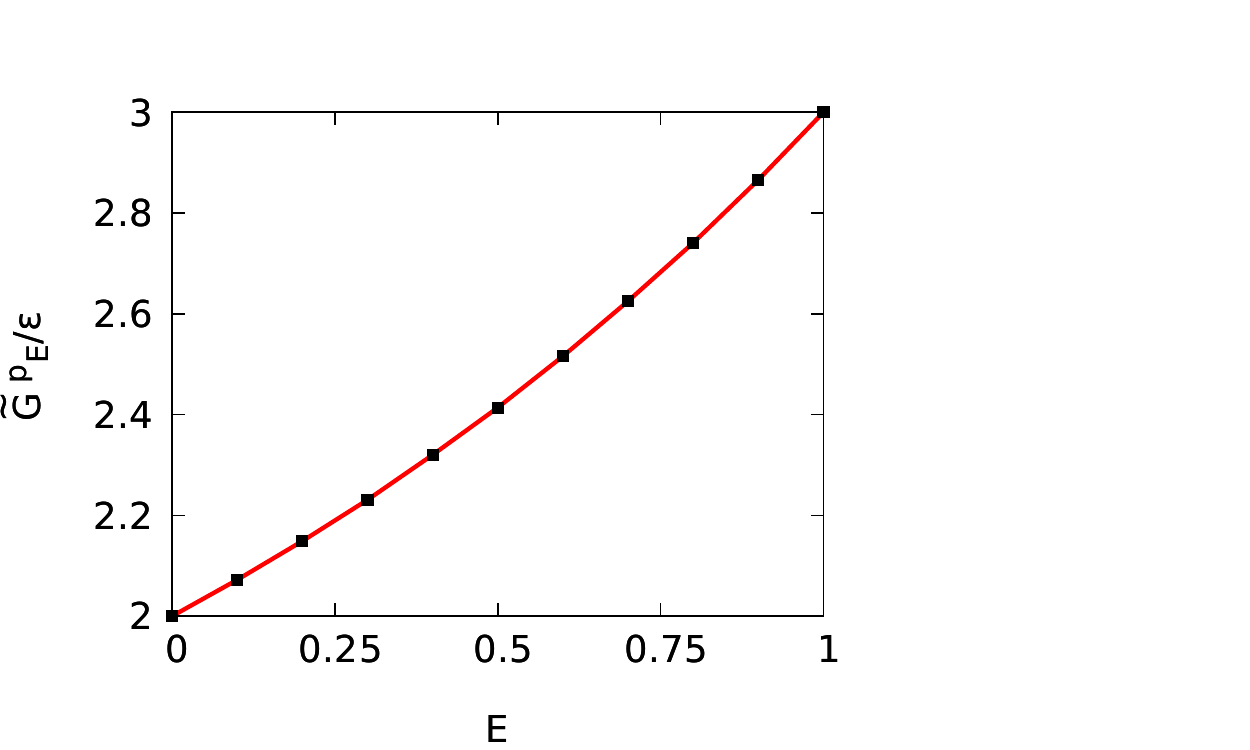}
\caption{Global work extractable from general two-qubit locally passive quantum battery states. The considerations are the same as in Fig. \ref{fig1}, except that the states can also be mixed, so that there are far more states available in the optimization procedure for a fixed value of $E$. The vertical axis is now to represent the quantity  denoted by $\widetilde{G}^p_E$, in units of $\epsilon$. Also, the plot is obtained via a numerical nonlinear optimization procedure.}
\label{fig2}
\end{figure}
We plot  $G_E^p$ vs $E$ in Fig \ref{fig1}. From the plot, we can conclude that the globally extractable work extraction increases with entanglement for states from which we can not extract any local work.

We numerically analyze the maximum work extraction for local passive states which may be mixed, via numerical nonlinear optimization. The runs are performed so that the maximum value of globally extractable work, which we now denote as $\widetilde{G}^p_E$, is correct up to the $3^{rd}$ decimal point. We present the graph in Fig. \ref{fig2}. Comparing Figs. \ref{fig1} and \ref{fig2}, we can see that global work extraction from a  mixed state is much higher than that for the pure state with the same entanglement for locally passive states. While $G^p_E$ and $\widetilde{G}^p_E$ are both concave upward as functions of $E$, although for pure states, the curvature is higher. For low values of entanglement, non-pure states provide far greater globally extractable work than pure states.
\section{Global work extraction from general battery states with fixed entanglement}
\label{sec4}
We now move over from locally passive states to general quantum states of two qubits, while still remaining with local Hamiltonians, and analyze the amount of work that can be extracted globally, from a state with a fixed value of entanglement. The case of local work extraction for general two-qubit states, and its difference with the globally extractable work, is considered in the succeeding section. 

We begin the analysis by considering pure states, so that the Hamiltonian and the states are of the form given in Eqs. \eqref{eq4} and \eqref{eq5}. In this case, the states which we take into consideration are not necessarily locally passive. Hence, the $c_i$'s satisfy only conditions (i) and (ii). The forms of $E$ and globally extractable work remain the same as in Eqs. \eqref{eq6} and \eqref{eq7}. Since, in this section, we are talking about global work extraction from  states that are not necessarily locally  passive, we will denote the maximum work extraction from a pure state by $W_{max}$ and the maximum work extraction from a set of pure states with fixed entanglement $E$ by $G_E$. Now, the coefficients of $\left(\epsilon^A+\epsilon^B\right)$ and $\left(\epsilon^A-\epsilon^B\right)$ have the same forms as in Eq. \eqref{eq7}, and both have the same constraints on them as given in the conditions (i) and (ii), and Eq. \eqref{eq6}, so that the maximum value that one of the coefficients can achieve, is the same for both of them. Now, since $\left(\epsilon^A+\epsilon^B\right) \geq \left(\epsilon^A-\epsilon^B\right)$, if we keep increasing the value of the coefficient of $\left(\epsilon^A+\epsilon^B\right)$ and keep decreasing the value of the coefficient of $\left(\epsilon^A-\epsilon^B\right)$ in a way such that the constraints remain satisfied, we can maximize $W_{max}$, keeping entanglement fixed. This maximum value of global work extraction is given by
\begin{equation}
 G_E=\left(\epsilon^A+\epsilon^B\right)\left(|c_0|^2-|c_3|^2+1\right). \label{eq9}
 \end{equation}
 Using Eq. \eqref{eq6} and condition (ii), we get the same solution for $|c_0|$ and $|c_3|$ as given in Eq. \eqref{eq8}. But in this case, $|c_0|$ may not be less than $|c_3|$, and we can see from Eq. \eqref{eq9} that we would get a higher amount of work extraction for $|c_0|\geq |c_3|$ in comparison to the case when $|c_0|\leq |c_3|$. Hence, in this case, we choose  $|c_0|=\frac{1}{2}\left(1+\sqrt{1-(2^E-1)^2}\right)$. Therefore,
\begin{equation}
  G_E=(\epsilon^A+\epsilon^B)\left(1+\sqrt{2^{E+1}-2^{2E}}\right). \label{eq13}
\end{equation}
The state, for which this maximum value is achieved, is given by
\begin{eqnarray}
\rho^{max}_{E}=\left[
\begin{matrix}
|c_0|^2&0&0&c_0c_3^* \\
0&0&0&0  \\
0&0&0&0  \\
c_3c_0^*&0&0&|c_3|^2 
\end{matrix}
\right], \label{eq10} 
\end{eqnarray}
where 
\begin{eqnarray}
|c_0|&=&\frac{1}{2}\left(1+\sqrt{1-(2^E-1)^2}\right) \nonumber \\
\text{and } |c_3|&=&\frac{1}{2}\left(1-\sqrt{1-(2^E-1)^2}\right). \label{eq12}   
\end{eqnarray}
\begin{figure}[h!] 
\includegraphics[scale=.8]{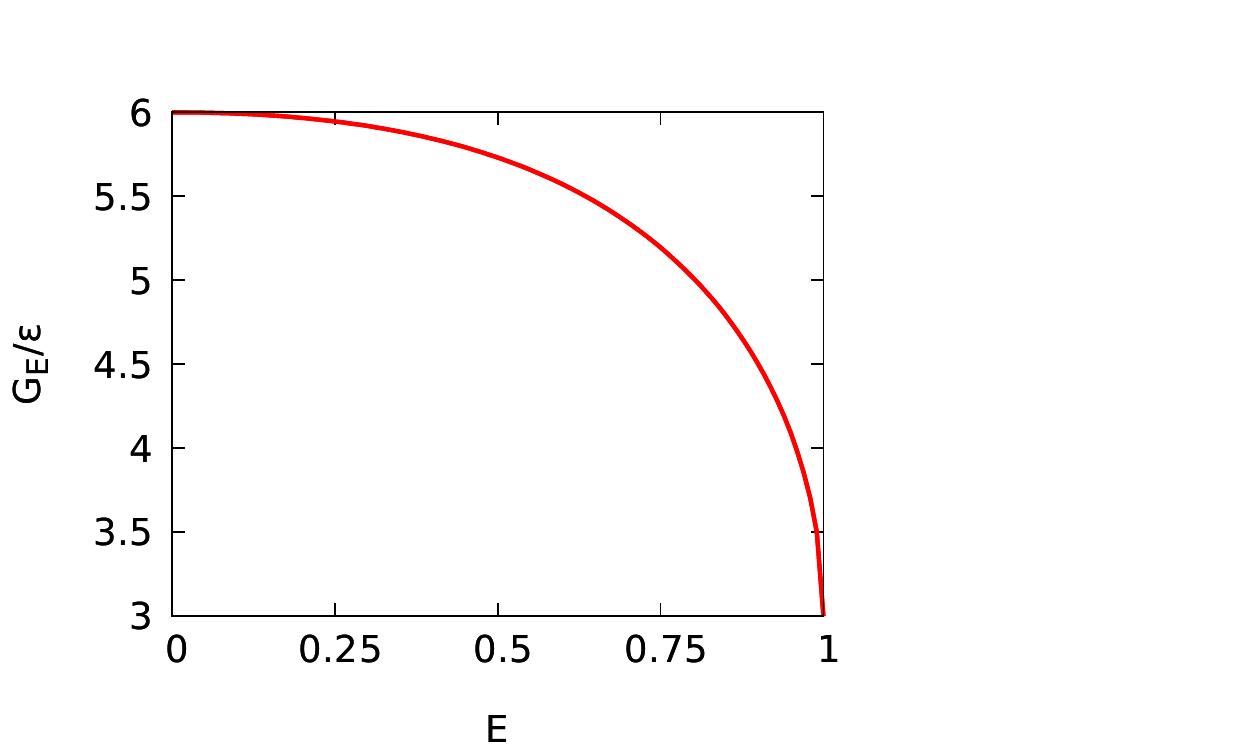}
\caption{How much work for given entanglement? We plot here the globally extractable work from an arbitrary pure quantum battery state with given entanglement. While the globally extractable work is represented on the vertical axis and denoted by $G_E$ (in units of $\epsilon$), the entanglement is represented on the horizontal axis and denoted by $E$. The vertical axis is dimensionless, while the horizontal one is in ebits. } 
\label{fig3}
\end{figure}
In Fig. \ref{fig3}, we plot this $G_E$ as a function of $E$. We can see that the two curves display respectively in Figs.  \ref{fig1} and \ref{fig3} have rather opposite natures. While $G^p_E$ is increasing with entanglement, $G_E$ is decreasing. And whereas $G^p_E$ is a concave function of entanglement, $G_E$ is convex. To get an understanding of this differing nature of the two curves we do an analysis at the beginning of the succeeding section.

\begin{figure}[h!] 
\includegraphics[scale=.8]{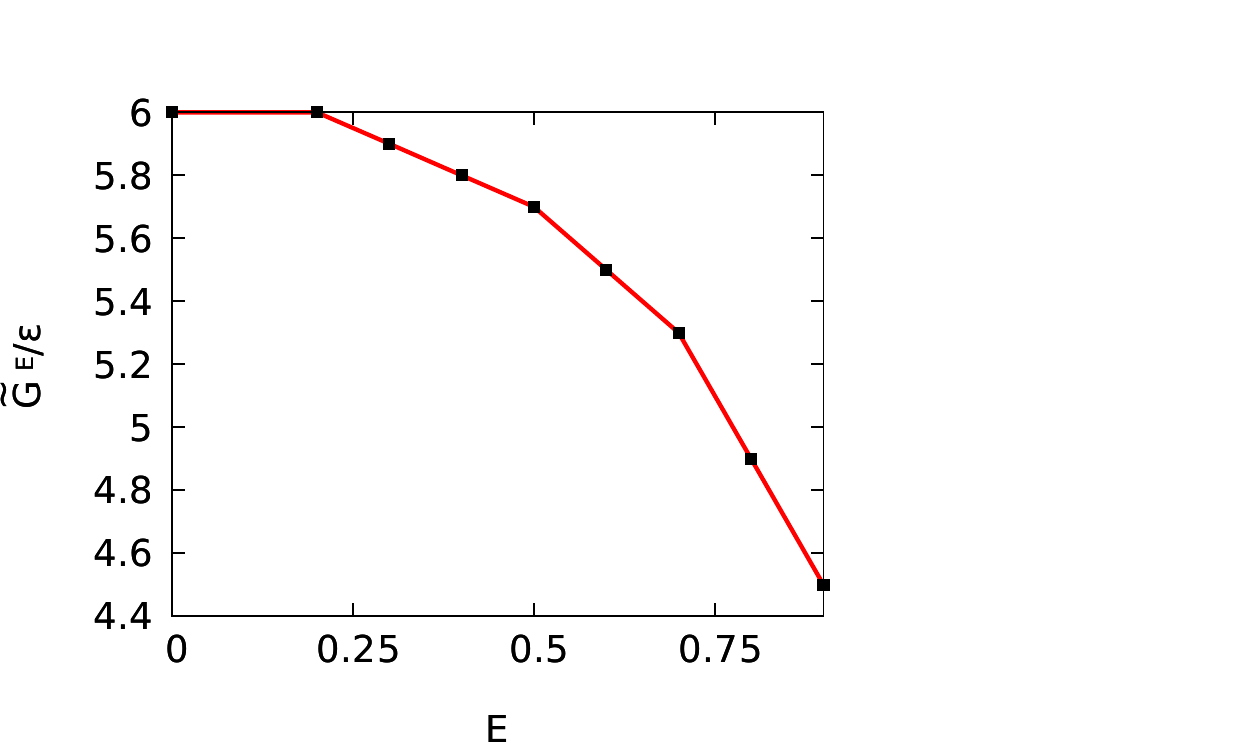}
\caption{Globally extractable work from general quantum battery states. The considerations are exactly the same as in Fig. \ref{fig3}, except that the battery state for a given entanglement, $E$, can be non-pure. The globally extractable work is denoted here by $\widetilde{G}_E$.} 
\label{fig5}
\end{figure}

We now move over to general states, and find the maximum amount of work that can be extracted by global unitaries. The analysis is again performed using the numerical nonlinear optimization procedure. The convergence is checked up to the first decimal point. In Fig. \ref{fig5}, we plot this maximum value as a function of the entanglement in the battery state. We observe that the behavior of the plot in Fig. \ref{fig3}, where the battery state was restricted to be pure, is similar to that in Fig. \ref{fig5}, where there is no such restriction.
\section{Work deficit for local extraction from quantum battery}
\label{sec5}
In this section, we will discuss about the quantity of advantage when performing global operations for extracting work in comparison to the case when local operations are performed. We will first determine the amount of work deficit for using local work extraction from states for which global work extraction is maximum among all pure states with fixed entanglement. Then we will find the difference between maximum amount of extractable work using global and local operations as a function of entanglement.

We have seen that the optimal amount of extractable work, among the set of all pure states with fixed entanglement $E$, is achievable for the state $\rho^{max}_E$, expressed in Eq. \eqref{eq10}. The locally passive state corresponding to this pure state,  is given by 
\begin{eqnarray}
\sigma_{\rho^{max}_E}^l=
\left[
\begin{matrix}
|c_3|^2&0&0&c_3c_0^* \\
0&0&0&0 \\
0&0&0&0 \\
c_0c_3^*&0&0&|c_0|^2 
\end{matrix}
\right]. \nonumber
\end{eqnarray}   
Here, $c_0$ and $c_3$ satisfy Eq. \eqref{eq12}.
Hence, using Eq. \eqref{eqB}, we get the locally extractable work from $\rho^{max}_E$ as  
\begin{equation}
\overline{L}_E=2(\epsilon^A+\epsilon^B)\sqrt{2^{E+1}-2^{2E}}. \nonumber \label{eq11}
\end{equation}
Therefore, the deficit in work extraction from $\rho_E^{max}$ because of restricting to local unitaries is
\begin{equation}
G_E-\overline{L}_E=\left(\epsilon^A+\epsilon^B\right)\left(1-\sqrt{2^{E+1}-2^{2E}}\right). \label{eq15}
\end{equation}
This is exactly equal to the amount of optimal global work extractable from pure locally passive states, for the same entanglement, i.e., $G^p_E$. We therefore have the following result.
\vspace{0.5cm}\\
\noindent \textbf{Theorem 3.} \emph{The difference in work extractions between the instances using global and local unitaries from the pure battery state providing maximal global work is equal to the amount of 
maximal global work available from the corresponding pure locally passive state having the same entanglement.}

Therefore,  we can conclude that the opposite features in Figs. \ref{fig1} and \ref{fig3} arose due to the fact that work was being extracted  from $\rho^{max}_E$ by using local unitary operations, where, since $\sigma^{l\text{ }max}_E$ is a locally passive state, the locally extractable work from $\sigma^{l\text{ }max}_E$ is zero.

\begin{figure}[h!]
\includegraphics[scale=.8]{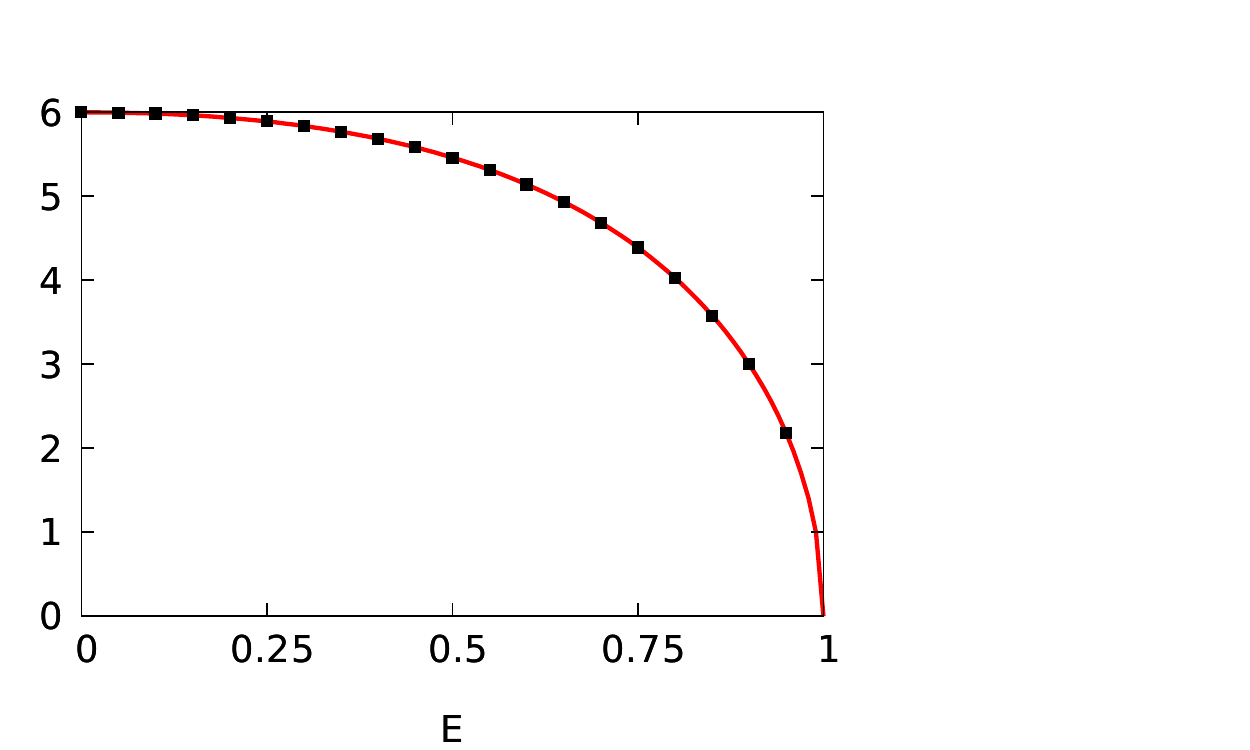}
\caption{Interplay between global and local work extraction for quantum batteries. We plot here the locally extractable work from pure states with fixed entanglement $E$, and locally extractable work from that pure state for which globally extractable work is maximum, with the same entanglement $E$. The optimal amount of locally extractable work from pure states, $L_E$, with fixed entanglement $E$, have been plotted using black squares, along the vertical axis, against $E$ along the horizontal axis. On the other hand, the amount of extractable work using local operations, $\overline{L}_E$, from a pure state with entanglement $E$ for which extractable work using global operations is maximum is plotted along the vertical axis using the red line. $L_E$ and $\overline{L}_E$ both are given in units of $\epsilon$, so that the vertical axis is dimensionless, while the horizontal axis is in ebits.} 
\label{fig6}
\end{figure}
We have found the maximum work extractable by global operations for pure states in the preceding section, and have obtained the result that the maximum extractable work, $G_E$ (given in Eq. \eqref{eq13}), is achievable for the state $\rho^{max}_E$ (given in Eq. \eqref{eq10}). Using the nonlinear numerical optimization procedure, we now find the maximum amount of work extractable from a pure state with entanglement, $E$, by using local operations, and denote it by $L_E$. We denote the state for which this locally extractable work is maximum by $\rho^{l \text{ }max}_E$. 
Surprisingly, we find that 
\begin{eqnarray}
\rho^{l \text{ }max}_E&=&\rho^{max}_E, \nonumber \\
 \text{and hence,   } {L}_E&=&\overline{L}_E. \label{eq14} 
\end{eqnarray}
This can be seen from Fig. \ref{fig6}, where we plot $\overline{L}_E$ with a line and $L_E$ with points. It can be noticed that all the points fall on the line. 
 Hence, we can state the following result. \vspace{0.5cm}\\
\textbf{Proposition 4.} \textit{The pure state for which the globally extractable work is maximum, among the set of all pure states with a fixed value of entanglement, also provides a maximum locally extractable work.} 

Using Eqs. \eqref{eq15} and \eqref{eq14}, we get the difference between maximum global work extraction and maximum local work extraction, as a function of entanglement, and is given by
\begin{equation}
G_E-L_E=(\epsilon^A+\epsilon^B)
\left(1-\sqrt{2^{E+1}}-2^{2E}\right). \nonumber
\end{equation}. 
\section{Conclusion}
\label{sec6}

We have analyzed the working of a shared quantum battery governed by local
Hamiltonians. An important attribute of a battery is its passive state,
that is, the state that disallows any energy extraction. A
characterization of the globally passive state was already known. Here we
have characterized the local passive state for an arbitrary quantum
battery with local Hamiltonians.

We subsequently restricted our attention to two-qubit systems. We found
the relation between the entanglement of a two-qubit locally passive
battery state with the amount of energy that can be globally extracted
from it. While the result is derived analytically for pure battery states,
the general case is determined via a nonlinear numerical optimization
procedure.

We then considered the question of global extraction of energy from a
general - not necessarily locally passive - two-qubit battery when the
governing Hamiltonian is local. We found that the difference between
global and local extraction of work from a pure battery state with a given
amount of entanglement is equal to the optimal global work extractable
from the corresponding locally passive battery state having the same
entanglement. We also showed that the pure battery state for which
globally extractable work attains a maximum, among the set of all pure
states with a fixed value of entanglement, also
provides the maximum locally extractable work.

\end{document}